\documentclass[aps,prd,twocolumn]{revtex4}

\usepackage{natbib}
\usepackage{dcolumn}
\usepackage{amssymb}

\input epsf
\def\plotone#1{\centering \leavevmode
\epsfxsize=0.80\columnwidth \epsfbox{#1}}
\def\plotonefull#1{\centering \leavevmode
\epsfxsize=1.00\columnwidth \epsfbox{#1}}
\def\plotonemedium#1{\centering \leavevmode
\epsfxsize=0.75\columnwidth \epsfbox{#1}}
\def\plotonenarrow#1{\centering \leavevmode
\epsfxsize=0.70\columnwidth \epsfbox{#1}}
\def\plotonetiny#1{\centering \leavevmode
\epsfxsize=0.55\columnwidth \epsfbox{#1}}
\def\plotonewide#1{\centering \leavevmode
\epsfxsize=1.35\columnwidth \epsfbox{#1}}

\begin{document}

\title{Peak Quasar Correlation Across the Sky, and in an Earth-Wide Bell Test}

\author{Eric Steinbring}
\email{Eric.Steinbring@nrc-cnrc.gc.ca}
\affiliation{National Research Council Canada, Herzberg Astronomy and Astrophysics, Victoria, BC V9E 2E7, Canada}

\date{\today}

\begin{abstract}
Viewing two astronomical sources at large enough distance and angular separation can assure, by light-travel-time arguments, the acausality of their emitted photons. Using such photons to set apparatus parameters in a laboratory-based quantum-mechanical experiment could ensure those switch settings are independent and fair, allowing a loophole-free test of Bell's inequality. Quasars are a natural choice for this task, yet at ultimate extent it involves their simultaneous photometry towards opposite directions on the sky, which is untried. Cosmic isotropy can be invoked to set limits there, leaving fairness intact for causal pairs, but with a testable consequence of asymmetric bias: mean brightness correlations found less flat in sky angle than random, more acutely so inside a horizon of $90^\circ$. Analysis of one dataset from the Gemini twin telescopes is presented, using over 14 years of archival broadband-optical images, serendipitously sampling thousands of quasars up to $180^\circ$ apart. These data reject a null result of no correlation with 97.7\% confidence, instead consistent with a $3\sigma$ residual signal of 0.21 mag peaked at $65^{\circ}\pm3^{\circ}$ separation. Possible confirmatory observations are pointed to along with the improved experimental protocol of an Earth-wide test.
\end{abstract}

\keywords{quasars, cosmology, techniques}

\maketitle

\section{Introduction}\label{introduction}

That quantum mechanics (QM) must be incomplete, allowing ``spooky" outcomes requiring either super-luminal signals or hidden variables, was famously contended by Einstein, Podolsky and Rosen \cite{Einstein1935}. Bell showed how correlations in a QM experiment could allow tests against such unsensed influences \cite{Bell1964}. Modern experiments routinely find QM is correct, having tightly and simultaneously restricted necessary conditions on measurements (e.g. \cite{Rosenfeld2017} and references therein) but not excluded a final possibility, by closing the so-called ``freedom-of-choice" loophole, eliminating experimenter interaction. One promising route is to set experimental parameters via photons from astronomical sources \citep{Friedman2013, Gallicchio2014}, requiring that interference between two settings had been orchestrated between distant sources and the Earth-based observer. Proof-of-concept QM tests using stars within the Milky Way have already been achieved \cite{Handsteiner2017, Li2018} forcing ``collusion" in the outcome back hundreds of years. And a recent observational development was the extension to quasars \citep{Wu2017, Leung2018, Rauch2018}: the combination of high redshift $z$ with large angular separation on the sky can place these entirely outside each others' light cone; for separations of $180^\circ$ this occurs when both sources have $z\geq3.65$. The independence of settings triggered by those photons is unspoiled by their communication, and absent correlated errors corrupting the signals prior to detection, forces any unexplained coincidence to be the result of unexpected synchronization between sources. Otherwise, the foundations of QM would indeed be in question.

The quasar-based QM experiment performed by Rauch et al. follows the methodology of Clauser \cite{Clauser1969} where an entangled pair of photons emitted from a central source are split between two optical arms and their polarizations are detected at receivers. While those entangled photons are in flight, a switching mechanism at each receiver (also co-located with a telescope) selects randomly between two polarization measurements at pre-fixed relative angles, chosen to test the maximum potentially observable difference from QM. In these first test runs that switch was set by the colour of the most recently detected quasar photon, using bright pairs viewed separately via two 4-m class telescopes from Observatario del Roque de los Muchachos on La Palma. One quasar pair was separated by $73^{\circ}$ on the sky, having $z=0.27$ and $3.91$, with another pair $84^{\circ}$ apart at $z=0.96$ and $3.91$. Their fluxes were sufficient to allow large sampling losses: a relative polarization measurement was retained only if both quasar photons arrived within the microseconds while the entangled photons were in flight, during runs lasting 12 and 17 minutes. Detailed analysis showed that neither the colours of the two quasars nor background noise against which they were detected (notably sky brightness) were correlated beyond measurement error, upholding QM against collusion between polarization settings.

Can a QM experiment utilizing antipodal, and truly acaussal, quasar pairs be performed? The primary hurdle to one as described above is that such sightlines are effectively impossible from a single location on the Earth. A spaceborne mission with sufficient field of regard might do so, possibly even via direct photon-counting of $\gamma$-rays or X-rays. It is notable, however, that despite decades of optical variability studies (e.g. \cite{MacLeod2010}) and extensive reverberation mapping having established the characteristic sizes of AGN disks on the order of light days across \cite{Mudd2018}, no monitoring campaign simultaneously viewing such sources outside each others' horizon is so far reported in the literature, at any wavelength. The difficulty from the ground is, of course, hindrance by the Earth. Radio telescopes do not gain a benefit in this regard, as dish elevations are well above the local horizon, regardless of Sun position. From the nightside, optical/near-infrared observatories are further restricted in workable separations, hemmed below about two airmasses. That incurs at least twice the zenith extinction for each, even under photometric skies, with similarly degraded seeing. Darkness reaches only $\sim 21~{\rm mag}~{\rm arcsec}^{-2}$ in the visible, which is relevant for two far-separated quasars, both typically fainter. What must be overcome is colour-discrimination of those, viewed independently and simultaneously from opposite hemispheres. This is undemonstrated, which although not obviating previous QM experiments, does set a bar to an irrefutable ground-based one.

Observational aspects of fair switching can be investigated without input to a QM experiment, so free from practical issues of communication between sites and internal apparatus. That is previously unexplored at quasar-separation angles greater than $90^{\circ}$, justifying the effort, as those fluxes (in ${\rm photons}~{\rm s}^{-1}$) would signal the switch settings in practice, and when their relative photometry (${\rm photons}~{\rm s}^{-1}~{\rm arcsec}^{-2}$) is ill constrained these choices may be subject to a hidden connection, unproved against intrinsic synchronization. Tolerances on a search for that while still closing all loopholes are up to one in four external-source switching photons before being spoiled, although plausibly as little as 14\% \cite{Friedman2019}. These are readily testable photometric limits, and a method of sensing average flux differences will be described here, for conditions where point-source photometry has sufficient sampling and sensitivity to reach the zeropoints of two identical instruments, within just a few percent error.

And at least one useful dataset to probe is already available at Gemini: on Maunakea in Hawaii ($19.82^{\circ}$N, $155.47^{\circ}$W, 4213 m) and on Cerro Pachon in Chile ($30.24^{\circ}$S, $70.74^{\circ}$W, 2722 m), that when each viewing a target near zenith, places those $95.5^{\circ}$ apart on the sky. These 8-m class telescopes have operated near-identical optical imagers continuously for more than 15 years, and a public archive eases aggregation of many serendipitous observations.  Although such data do not, in themselves, constitute a QM experiment, they may provide a baseline in devising a future one: at over 10600 km apart, no collusion is possible on timescales less than this distance divided by the speed of light or $l/c\approx0.04~{\rm s}$, which in the restframe at $z=4$ corresponds to $0.2$ s.

The next section describes how the geometry of an experiment allowing simultaneous photometry of widely-separated quasar pairs restricts their best relative signal-to-noise ratio, and so could hide an underlying intrinsic synchronization, if present. This follows from recognizing an asymmetry in reflection about their averaged correlation, while satisfying both acausality (minimally, just for each antipodal pair with sufficient redshift) and cosmic isotropy (globally, among all sources) simultaneously. Thus, demanding fair switching photons can allow a potential form of correlated fluxes with angular separation: suppressed outside the horizon of any single site, retaining a true loophole-free test, if counterbalanced by detectable, possibly QM-mimicing residual peaks; especially prominent at $64.6^{\circ}$. In an attempt to rule this situation out a ``virtual test" is suggested with sources chosen in a randomized way to avoid bias and sky conditions sampled sufficiently to remove their influence. Following that, the available Gemini dataset is described, which consists of photometry in ${\rm g}$, ${\rm r}$, ${\rm i}$ and ${\rm z}$ filter bands for thousands of quasars with redshifts $0.1<z<6$, sensitive down to 23 mag. The final sample comprises roughly 2 million observational pairs, which in their aggregate (0.25-mag $1\sigma$ uncertainty within 6-degree-wide sampling bins) is sufficient to show a biased difference in brightness relative to object separation, reaching a peak signal consistent with the model and against a null hypothesis of flat. Although this suggests an intriguing connection between those quasar fluxes, it is not in conflict with either causality or previous QM results. Follow-up targets of interest are provided. Discussion concludes on prospects in the era of 30-m telescopes situated in both hemispheres, and reaching necessary photometric accuracy to exclude both any intrinsic correlation and local noise sources in closing the last observational loophole.

\section{Quantifying Quasar Independence}

The intent is to quantify a possible lack of randomness in external source fluxes relative to local noise at the receiver telescopes, not details internal to the apparatus, and so a basic description of a QM experiment is sufficient to illustrate how this can be connected to angle-setting independence. Generically, quantum theory demands that entangled photon pairs must be found in opposite polarization states; if one is found with horizontal polarization, the state of its entangled twin will always be found vertical. (In the original theoretic treatment, these were the spin states of entangled electrons: up or down.) Importantly, any real experiment cannot measure both states in one direction simultaneously, as this requires a setting change. For example, polarimetry necessitates a discriminator, such as a polarizer or the rotation of a waveplate. A choice must be made as to which polarization angle (or arm) to sample. Detecting the state of one entangled photon instantly collapses the wavefunction of both subject to shared uncertainty, with a probability density $q$ which depends on the angle $\theta$ between waveplate settings. Those states must be anti-correlated when co-incident (there are exactly two possibilities) and preservation of equal average probabilities of both states implies no net correlation, and so a functional form for normalized correlation of $-\cos{\theta}$, crossing at $\pi/2$. But when the experimenters' ability to freely choose settings avoids making them complicit in the outcome, this takes on instead the form \citep{Bell1964}: $$p(\theta) = |\cos{2\theta} - \cos{\theta}| + \cos{\theta}, \eqno(1)$$ which exceeds unity at all lesser angles. Detecting unequal or ``excess" correlation, above equality and beyond what truly random sampling predicts, would reveal a fundamental fault in QM.

\subsection{Potential Correlations Across the Sky}

In an experiment with fair switching there is no influence on results caused by synchronization between external sources. These are random and neutral, without bias in their triggering property: any red or blue photon (within suitably defined passbands, corrected for relative colour) from one source will correspond just as likely as not with receiving the opposite from the other. This is connected to the probability of a measureable difference in outcomes, given by the trace-distance metric \citep{Friedman2019} $$P={1\over{2}}+{D(p,q)\over{4}}, \eqno(2)$$ as this would be 50:50, that is $D=|p-q|=0$ or $P=50\%$ for unbiased switches. But synchronization is possible for causally-connected sources, influencing the outcome, as when external differences in red versus blue are neutral for only one or less of every 4 photons, so non-neutral for every 3 or more. And some maximal correlation $M$ could be reached at which an effect on $D$ cannot be hidden, before being necessarily spoiled by full mimicry of QM, at 100\%.  If so, there exists two perfectly synchronized sources, giving the opposite trigger to every photon arriving at either receiver, and (absent measurement error) {\it always} setting their polarizer angles, $\theta$, to correspond. 

Where antipodal, experimenters at the two telescopes (in mirrored, back-to-back configuration, even if remote) could then exploit an anisotropy, pointing in one particular spatial orientation for which they need never know the other's settings, just a favoured sky position in equatorial right ascension $\rho$ and declination $\delta$. In this case $D=2$, and it depends {\it solely} on the angle $\phi$ between sources: $q(\rho, \delta)\leftrightarrow q(\phi)$ and similar for $p$ at $\phi=\pi$, independent of $\theta$. When influence via $\phi$ merely dominates any from $\theta$, raised to a (peak-to-valley) distance \footnote{By adding up the 12 unique terms thus formed with $\pm M$, $(D/2)^2=(p-q+M)(p+q-M)+(q-p+M)(q+p+M) +$ ..., all the cross terms cancel.} $$D(\phi) = \sqrt{M^2+p^2(\phi)-q^2(\phi)} ~; ~0.75 \leq M \leq 1, \eqno(3)$$ locating a ``special" non-neutral pair (usually synced) could still trigger results to correspond more often than not. This motivates contriving a relaxed metric $$D_R/2 \equiv 3[D(\phi) - R]/4, \eqno(4)$$ anticipating how $R$ could manifest as a shared ``residual" correlation through cross-comparison of pairs, avoiding the stark, singular anisotropic-antipode case.

\begin{figure}
\plotonenarrow{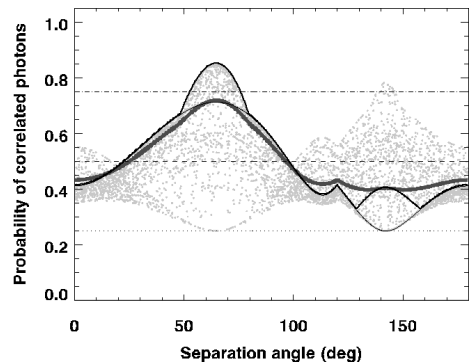}
\caption{Probability by source separation of non-neutral external switching photons in the imagined full-sky search described in the text (thin black curve: idealized case of $M=1$, $R=0$; thick black curve: $M=0.87, R=0.66$) against global neutrality ($\bar{P}=50\%$: dashed line) and an ensemble of $2\times 10^6$ random residuals (grey points; a subsample of 10000) preserving insufficient correlation to spoil a QM-experiment when antipodal (thick grey curve) and within thresholds of 25\% (dotted) and 75\% (dot-dashed) otherwise. Sources above the last would belong to special, synchronization-biased cases.}\label{figure_probability}
\end{figure}

Nature might provide such special sources, and an expectation of cosmic isotropy puts limits on finding those, as one could seek, via their average correlations, experimental pairings more likely to provide synchronized photons. To find them, take a census of all pairings $$\bar{P}=\int\displaylimits_{\rm R.A.}d{\rho}\int\displaylimits_{\rm Dec.}d{\delta}{P}={1\over{2}} + {\overline{D_R}\over{2}} ~~~~~~~~~~\eqno(5)$$$$~~~~~~~~~~~~~~~~~~~~~~~~~~~~~~~~~~~~~={1\over{2}} + {3\over{4\pi}}\int\displaylimits_0^{\pi} d\phi[D(\phi) - R],$$ tallying by each $\delta$ in polar coordinates about azimuth $\rho$, or equivalently the projected great-circle angular distance $\phi$ between any two sources. Isotropy demands $\bar{P}=50\%$ (all sky directions equal, with no preferred orientation) thus, by inspection, the intregral in equation 5 must be within the lower and upper possible limits of equation 4 for $R=\overline{|D_R|}=0$, presenting the desired tool: $$D_0\equiv(|D_R|-R)/2~~~~~~~~~~~~~~~~~~~~~~~~~~~~~~~~~~~~~~~~~~~~~~~~~~~\eqno(6)$$$$~~~~~~~={3\over{4}}\sqrt{1 + p^2(\phi) - q^2(\phi)}-0.88300... ~;~ M=1, R=0, $$ the constant arrived at by numerical integration of the left-hand term. Adopting instead $M=\sqrt{3/4}\approx0.87$, fairness in equation 2 is still not violated at antipodes, but continuing to bound $D$ by $-2R$ ($P=25\%$; resulting in 1 of 4 switching photons corrupted) everywhere, it has an upper limit $|D_R|+R$, defining an allowed envelope: $$P(\phi)={1\over{2}} + {D_0\over{2}}, ~~~~~~~~\eqno(7)$$ where $|D_R|/2 \leq R$, counterbalanced above by $$P(\phi)={1\over{2}} + 2 D_0 - {3\over{4}} R, \eqno(8)$$ that is, a probability distribution with inflections at $R$ (and a cusp at $4R/7$) revealing those pairings that offer more-often-than-not synchronized photons: above the $P=75\%$ threshold, non-neutral over 25\% of the time.  Most tellingly, equation 8 can balance for $R\approx 0.66$, thus reducing equation 8 to $2D_0$. This gives a global maximum, occuring at $\phi=64.6^\circ$ (a peak set by twice the observable difference; thick black curve in Figure 1) which is notably different from that of $p$, at $60^\circ$.

In effect, $D_0$ is akin to an experimental zeropoint, gauging a minimum synchronization which avoids QM-mimicry at source antipodes. This limit is refinable via cross correlation, with sufficient sampling.  Why averaging over many trials can facilitate its detection and show bias is illustrated in Figure~\ref{figure_probability}, the results of a simple simulation: $2\times 10^6$ realizations of $P$ for $M=0.87$ by randomly selecting residuals from a Gaussian distribution of width $R=0.66$ centered on $0.50$ (grey points). Sometimes an excess correlation for a given source-separation is found, other times not. On average, it is apparent: the mean of trials at each angular separation indicated with a thick grey curve, which (overall) differs from the expected mean of $\bar{P}=50\%$ by an increment just under 1.8\%, here subtracted - essentially incuring a small anisotropy. In this case, a fraction just over 4\% of sufficiently-synchronized cases would rise above the critical, 75\%-detection level, indicated by those instances above that bar, within the peaks. Put another way: if correct, in a large random sample of causally-linked sources uniformly distributed over the sky there could be about 20 out of 10000 sensed with 10 such biased triggers each (with more expected near $65^{\circ}$, as it is maximally bounded by equation 8 there) and no more than one of those beyond $90^\circ$ (the 20 points at or above the threshold in the right-hand peak).

Three key features of this potential correlation appear relevant to an Earth-wide experiment: First, finding synchronized photons seemingly sufficient to trigger a QM-mimicing result (just over 25\% more likely than neutral) for some sources separated by angles near its peak at $65^{\circ}$ would not be in conflict with previous experiments, which utilized quasars closer to $82^{\circ}$ separation, where expected influence passes below the limit of corruption (less than 14\% biased). Second, despite preserving isotropy it is distinctly bimodal and asymmetric about $90^{\circ}$ in separation, not merely sinusoidal, crossing the mean beyond $96^{\circ}$ and the median at $103^{\circ}$. This distinct attribute is required to preserve an average of neutral correlation of an ensemble across the whole sky: net correlation (after subtraction of the mean) is counterbalanced at larger angles. Past a distinctive ``kink" at $120^{\circ}$ they can undergo a further inflection, for angles between about $128^{\circ}$ and $159^{\circ}$, and so remain correlated up to angles impossible to reach from any single site. And third: in this form correlation is indeed suppressed (mean probability of synchronization not straying from fairness by more than 7\%) at $180^{\circ}$, serving to confirm the applicability of an Earth-wide test using antipodal sources, where those can be (minimally) acausal, as required for a true loophole-free test.

So far this is a generic, essentially geometric argument; apart from requiring isotropy while assuming synchronized causal sources do exist, no conditions have been placed on the properties of those, such as flux or temporal behaviour on any timescale. For example, if experimenters are ``unlucky" and two happen to be in sync and visible within just one night every few years, they are not likely to be discovered in practice. Perhaps those can defeat efforts given sufficient mutual incoherence, unknown intervening effects along photon trajectories, or where at insufficient distances to avoid these being effectively internal to the experiment. As stated, the proposed metric simply gauges how correlated external sources may be against the global average in a measurable property. So conceivably this is the likelihood of a particular relative polarization angle being detected, hinting at a more direct physical implication related to Bell's theorem, although that is not pursued further here. Photon colour corresponds to the switching mechanism in previous experiments, and only fluxes in defined passbands are available in this study, so the approach will be to probe whether such correlations with angular separation on the sky can be found.

\subsection{Angular Dependence of Noise}

In a true loophole-free QM test the switching photons effectively operate the apparatus, automating setting changes photon-by-photon between settings by pre-defined selection criteria. Any criterion can only ever be a relative flux measurement over a suitably defined passband and time period. So an issue would arise if the colour of sensed switching photons at the telescope were dominated by local noise. Either the sources or switching mechanisms could retain hidden correlation. Admittedly, unless strong, that may not predict the colour of the next photon to arrive, so not exploitable to mimic QM behaviour. But even if weak, experimental results are immediately connected through the form of $P$ to the allowable relative noise between telescopes. That correlation will be unavoidable from the ground, as seeing, sky background and extinction are well known to depend on airmass, and will inject some cosine dependence on errors with viewing angle $\phi$, which to be successful must be excluded against that of the source fluxes.

\begin{figure}
\plotonemedium{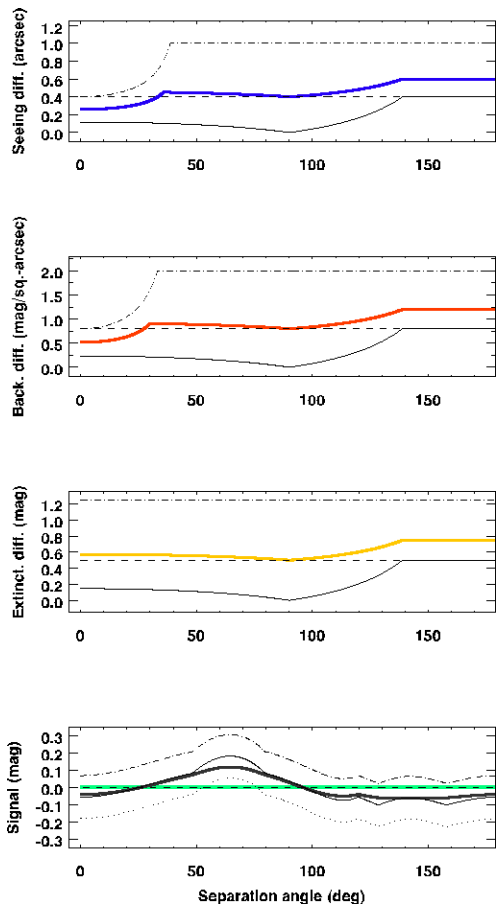}
\caption{Model differences in seeing (blue), sky brightness (red), and extinction (yellow) between two positions on the sky up to $180^{\circ}$ apart, viewed from two different sites, here plotted for Gemini. Instantaneous lower limits (thin black curves) and upper limits (dot-dashed) are indicated, giving those means of coincident observations and long-term medians (dashed lines) for combining views from each single site. Below is the long-term difference of two quasars selected from a brightness distribution of width 1.0 mag, if their fluxes were correlated as per probability $P$ at $S/N=3$ (thin black curve; $5\times$: dotted, dot-dashed) or random residuals (thick, grey) against their averaged instrumental errors (thick, green).}\label{figure_models}
\end{figure}

Consider identical instruments at two sites with a geographic separation of $90^\circ$, that is, two sources viewed simultaneously at zenith are at right angles. (This is essentially the case for Gemini North and South: $95.5^\circ$ apart.) Under clear skies, atmospheric extinction increases linearly with airmass, inversely with zenith distance ($1/Z$), so to a target under an airmass of 2, the difference between it and another $\phi$ degrees away, is at its extreme $$E(\phi) = \Big{|} 2 - {\sqrt{2}\over{\cos{(\phi/2)}}} \Big{|}, \eqno(9)$$ or $$A = \alpha E, \eqno(10)$$ where $\alpha$ is half the median extinction in magnitudes. A similar relationship can be found in sky brightness, $$B = \beta E, \eqno(11)$$ and image quality $$C = \gamma E_\gamma, \eqno(12)$$ where $\beta$ is in units of ${\rm mag}~{\rm arcsec}^{-2}$, and $\gamma$ is in arcseconds.  The functional form of $E$ has a zenith distance more like a standard expression \cite{Kasten1965}, which is $Z=\cos{(\phi/2)} + 0.150 \times (93.885 - \phi/2)^{-1.253}$, and for seeing ($E=E_\gamma$) is perhaps better modeled with a weaker $Z^{0.6}$ power. And while one site is still enjoying better weather, the other might continue imaging only under 80\%-ile conditions, a relative factor $2.5\times$ poorer than the median. Those limits are shown in Figure~\ref{figure_models}, represented by a thin line and a dot-dashed line respectively. Later it will be shown that for Gemini $\alpha\approx 0.25~{\rm mag}$, $\beta\approx 0.40~{\rm mag}~{\rm arcsec}^{-2}$, and $\gamma\approx 0.38~{\rm arcsec}$; so those values are adopted in Figure~\ref{figure_models}. Their averages are shown as thick curves.

\subsection{Observationally Detectable Signal}

The requirement on how uncorrelated sources must be to avoid this bias, if real, can follow by working backwards from the needs of setting independence: finding sufficient flux difference beyond observational noise for two randomly-selected sources to be confident that switching based on those was {\it not} random. Although quasars are known to fluctuate on timescales of days to many years, and likely do so on timescales as short as the QM experiment, as a group they have well-studied optical brightness distributions. So if those were sampled in a perfectly unbiased way from a distribution of width $\omega$ their maximal long-term difference as amplified by equation 2, normalized by peak $\hat{P}$ to its mean is $$S = \omega P (1-\hat{P})/\bar{P} \approx 0.27 \times \omega P, \eqno(13)$$ in magnitudes. The signal amounts to an excess brightness relative to flatness with separation angle, and so the problem becomes one of determining how many quasars to sample at random, for how long, and how accurately to overcome uncertainty in flux measurements, which at minimum will be restricted by the instrumental error in relative flux difference, and from the ground is likely further impacted by variation of sky brightness, seeing, and atmospheric extinction on similar timescales.

The signal-to-noise ratio of detectable enhancement in flux differences over observational noise for $n$ samples thus has a form $$S/N \propto {\omega P \over{(\alpha E + \beta \gamma^2 E^3 + \epsilon)/\sqrt{n}} + \zeta}, \eqno(14)$$ relative to an ideal $\overline{S/N}=1/0.27$ limit, where $\epsilon$ is the photometric uncertainty and $\zeta$ is the bandpass zeropoint error, both in magnitudes. Note that, as they are maximal, $A+B+C$ are not added in quadrature, and fall off as the square-root of the number of samples, as does photometric uncertainty. Binning the data can enhance $S/N$ only until it reaches the zeropoint accuracy. The width of the quasar-brightness distribution in the optical is roughly $1.0$ mag, and zeropoints not typically better known than to 2\%, so the maximum detectable effect in a randomly selected pair, even to a space-based observation, is limited to about $\omega / \zeta \lesssim 50$.  

From the ground, and extended to a long-term average, the instrumental errors average out across the sky over many samples. Thus $E_\gamma=E=1$, which implies that (sampled over all angles) the effect slowly grows to $S/N\approx2.9$ again by $n=10000$ ($3.7$ at $2\times10^6$, if those were truly independent) with photometric uncertainty fixed at $0.25~{\rm mag}$. This suggests that to achieve similar constraints either a single pair with minimum observational noise or a larger, noisier (but binned) sample across the full sky may present comparable ways to detect the influence of $P$. It is the latter method which is adopted here: obtaining good relative photometry of an unbiased quasar sample through many years, and looking for a relative dependence with angular separation. Relative measurements of a given angle should still be dominated by their shared photometric uncertainty until sufficient samples beat that down below twice the zeropoint error, which for $0.25~{\rm mag}$ occurs at $n=10$. For reference, 10 Gemini FoVs ($5~{\rm arcmin}$ across) spans roughly one degree. Averaged over the sky, a detection minimum, at $S/N=3$, is shown in Figure~\ref{figure_models}. The form of this curve suggests a further constraint: the main peak and inflections are about $30^\circ$ across (minima to minima), crtically spanned by 5 samples; two per side would be optimal bins about $6^\circ$ wide, so the cusp at $\phi=120^{\circ}$ is minimally spanned by 3, and still marginally sufficient to discern a difference between a peak at $64.6^{\circ}$ from $60^{\circ}$. 

\section{Sample}\label{sample}

Archival Gemini Multi-Object Spectrograph (GMOS) images were searched for all instances of a known quasar falling into the field, starting from the beginning of regular GMOS operations to the beginning of 2016, spanning 14.5 years. A difficulty with a direct search is that the file header information will include a target name selected by the observer, which may not necessarily correspond to any catalog. Also, this would exclude cases where an object happened to fall on the detector during the observation of another, defined target. So instead the Million Quasars Catalog (MILLIQUAS), Version 4.8 of 22 June 2016 was cross-correlated with the full Canadian Astronomy Data Center (CADC) archive of science frames obtained with GMOS North and South. The MILLIQUAS is a compendium of published catalogs, primarily the Sloan Digital Sky Survey (SDSS) providing a redshift plus optical magnitudes (Blue: B, V or ${\rm g}$; and Red: ${\rm r}$ or ${\rm i}$) for each object. No restriction on type was made, including gravitational lenses, but the likelihood of misidentifications is considered small.

The GMOS field of view (FoV) is approximately $5 \times 5~{\rm arcmin}^2$, with 0.075 arcsec pixels. Using the Common Archive Observation Model (CAOM) Table Access Protocol (TAP) web service and custom Python scripts, a search inside a 5 arcmin radius for each object was conducted, which produced 28374 cases where an $0.1 < z < 6$ object was within a GMOS field in ${\rm g}$, ${\rm r}$, ${\rm i}$, or ${\rm z}$ (similar to the SDSS filters) through the full range of right ascension, and between $-79^{\circ}$ and $+82^{\circ}$ declination; deleting those corrupted or otherwise unusable yielding 20514 objects (${\rm g}$: 4691, ${\rm r}$: 6709, ${\rm i}$: 6776, ${\rm z}$: 2338), and a typical exposure time of 150 s. Gemini grades each science frame by clear-sky-fraction bins from CC20 (best 20\%-ile) to CCAny (all conditions). This represents 11483 unique (although potentially repeated) objects having a mean redshift of 1.48, no noticeable dependence on sky position, and obtained under all conditions under which the telescopes were operational.

Finally, a selection was made to ensure good data quality in each observation. Every object frame was searched for a comparison star from the Fourth U.S. Naval Observatory CCD Astrograph Survey (UCAC4) to serve as a photometric calibration, with published SDSS ${\rm r}$ magnitude. There were 16774 frames that had such a suitable, unsaturated star. This also provided each an image quality criterion; image Point-Spread Function (PSF) Full-Width-at-Half Maximum (FWHM) was extracted for each star. The FWHM of the object (also confirmed to be unsaturated) was checked against this; in just a few cases where it occurred, the smaller of the two was taken.  A lower FWHM limit of 0.20 arcsec was applied to ensure that only real images were obtained, without artifacts, and so it reflects seeing conditions. After this, only 9317 objects remained after culling to cloud-cover conditions better than CC80, or usable, with less than 2 mag of extinction.

\begin{figure}
\plotonenarrow{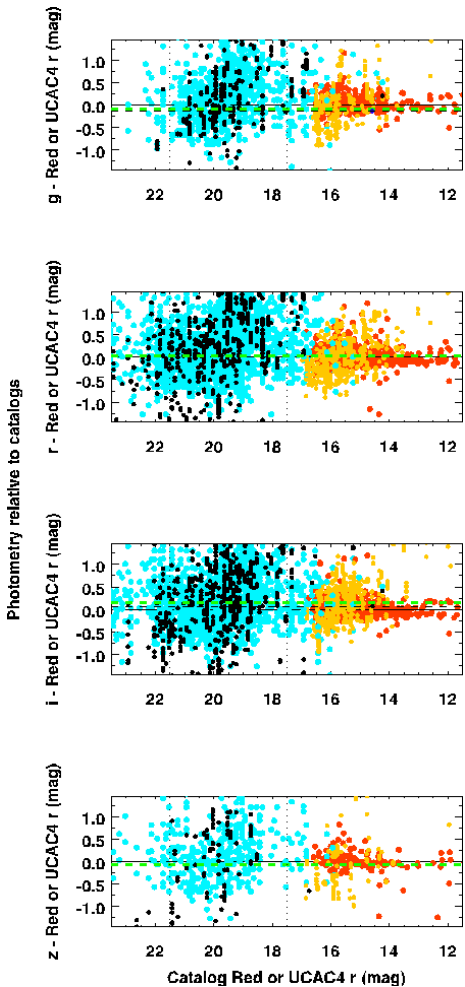}
\caption{Differences between observed and catalog magnitudes for objects (black) and comparisons (yellow) in the four filters studied. The relative colours of those against MILLIQUAS Blue (light blue) and UCAC4 ${\rm i}$ (red) are also shown; comparison colours (thick-dashed green) are neutral, as are the objects (dashed black). Those last have been limited to magnitude cutoffs (vertical dotted lines), serving to avoid a sample colour bias; if perfectly Gaussian each would be zero.}\label{figure_by_filter}
\end{figure}

\begin{figure}
\plotonemedium{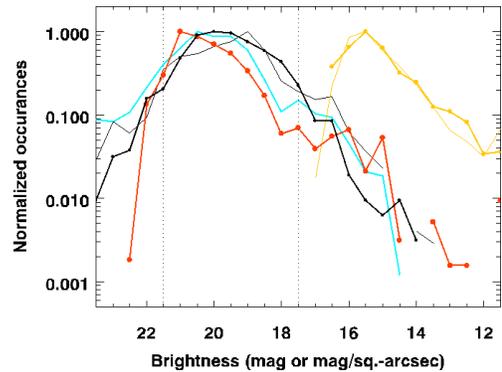}
\caption{Histograms of sky background (red curve), object (black), and comparison star (yellow) magnitudes, normalized to peak occurrences. Observed magnitudes are comparable to catalog MILLIQUAS Blue (light blue) and Red (thin black) values as are comparisons and catalog UCAC4 r (thin yellow).}\label{figure_magnitude_histogram}
\end{figure}

\section{Photometry}\label{photometry}

Synthetic aperture photometry was carried out on the full sample (objects and comparison stars) using a 4 arcsec diameter aperture throughout. Roughly 20 ${\rm arcsec}^2$ postage stamps sections were downloaded from the database. The positional accuracy of the frames was found by inspection to not always be better than 2 arcseconds. During the FWHM measuring step, a centroiding algorithm located the central pixel position and then sub-pixel shifted the aperture prior to obtaining the flux. Median sky backgrounds for each frame were subtracted after applying the appropriate detector gains and filter zeropoints, as published on the Gemini webpages. The detectors, and their layout in the focal plane changed at certain times, either 4 or 6 chips across the FoV with separate amplifiers and gaps between them, and each detector has a slightly different amplifier response. On average these are $28.11$, $28.31$, $28.16$ and $27.17$ mag in ${\rm g}$, ${\rm r}$, ${\rm i}$ and ${\rm z}$, which fluctuated over that time with deviations of 0.12, 0.02, 0.12 and 0.22 mag respectively.

All resultant photometry was corrected for atmospheric extinction using the calibration stars, where each of those was relative to its UCAC4 ${\rm r}$ brightness and a mean filter correction; results are shown in Figure~\ref{figure_by_filter}. This was calculated using the 237 observations with complete photometry in MILLIQUAS catalogued B, V, ${\rm r}$ and ${\rm i}$ magnitudes, which allows the calculation of a mean sample colour shift to B, V, ${\rm r}$, and ${\rm i}$ of 0.84, 0.49, 0.34, and 0.19 mag: g was interpolated as 0.66 mag and ${\rm z}$ extrapolated to 0.18 mag. Of these, 124 were cases of objects included in the UCAC4 catalog, all brighter than 17.5 mag. Correcting stellar colours so that the average sky background difference across the full sample is zero in ${\rm r}$ yields corrections of 0.50, 0.00, -0.27, and -0.22 mag in ${\rm g}$, ${\rm r}$, ${\rm i}$, and ${\rm z}$ filters respectively; although less of a concern in ${\rm r}$ and ${\rm i}$ data, the mean values of the resultant sample in all four filters are very close to neutral. The objects were also shifted by the same mean filter correction to ${\rm r}$ and uniformly taken as a differential from their catalog magnitudes. In this way, all photometry is relative to the sample average of ${\rm r}=20.0$ mag; photometry is sky-background-limited, with some down to the expected $5\sigma$ point-source limit of 23 mag: see Figure~\ref{figure_magnitude_histogram}. A lower object cutoff at 21.5 mag excludes those fainter than the mean sky surface brightness, avoiding a colour bias.

\begin{figure}
\plotonenarrow{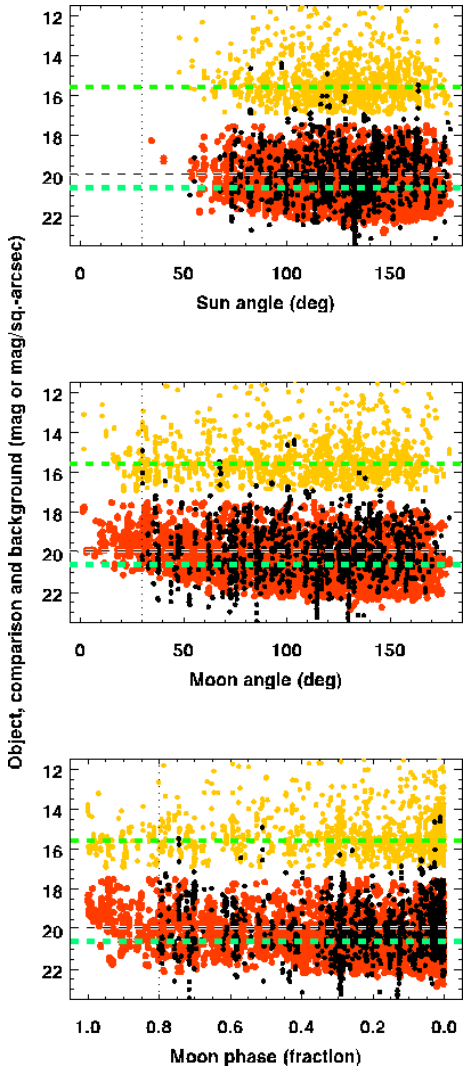}
\caption{Object, comparison and sky background with angle to Sun and Moon, and Moon phase; a proximity limit of $30^\circ$ and 80\% full Moon (horizontal dotted lines) ensures skies were suitably dark; medians shown as dashed horizontal lines, against comparisons and sky background (thick, green).}\label{figure_with_sun_and_moon}
\end{figure}

\begin{figure}
\plotone{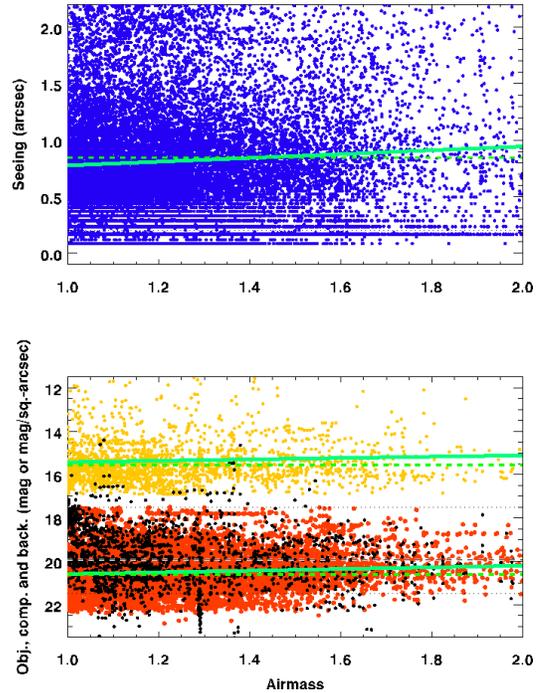}
\caption{All object magnitudes obtained (filled black circles) by airmass, with the UCAC4 ${\rm r}$-band magnitudes for comparison stars in each field (yellow) and their background sky brightnesses (red); sample limits have been applied (horizontal dotted lines). Above: image quality estimates for each frame (blue); data are spurious below a cutoff at 0.20 arcsec (horizontal dotted line). Medians and linear-least-square fits are shown for comparison (thick, green).}\label{figure_object_by_airmass}
\end{figure}

The weak influence from sky brightness is illustrated in Figure~\ref{figure_with_sun_and_moon}. Only mild angle and sky-brightness restrictions were employed, with a uniform upper cutoff of $17.5~{\rm mag}~{\rm arcsec}^{-2}$. To meet this, objects must have been at least $30^\circ$ from the Sun or Moon, during phases less than 80\% full (see Figure~\ref{figure_with_sun_and_moon}). This agrees well with expectations from the linear model, seen in Figure~\ref{figure_object_by_airmass}. Linear least-squares fits to extinction (comparison magnitudes), sky brightness and seeing (image quality) are shown in thick green (medians: dashed), giving $\alpha=0.25$ mag, $\beta=0.40 ~{\rm mag}~{\rm arcsec}^{-2}$, and $\gamma=0.38$ arcsec.

\section{Analysis and Results}\label{results}

Correlated observations were based on the sky position and UTC time recorded in each frame header.  An Interactive Data Language (IDL) code was written to perform this step, with the following prescription. For each observation, all earlier ones were searched to find exposure duration overlapped by some fraction with the current one. A positive fraction means that there is overlap. Perfect overlap, or perfect coincidence, would be a fraction of unity, and all less - down to zero - indicates that for equal fluxes this fraction of photons in the exposure would overlap with those taken in the other: over 50\% is considered ``coincident."  A negative fraction means that there is no overlap in the exposure durations, however, it is still useful in characterizing the temporal aspects of the sample. Observations were considered ``correlated"  if they fell within a given temporal window.  For example, an initial check was to find if another observation occurred during the same observational ``day" at the combined Gemini telescopes, which varies during the year, but is about 11 hours plus the timezone difference, or 17 hours. The largest possible window was to look back 14.5 years, to the beginning of records.

\begin{figure}
\plotonefull{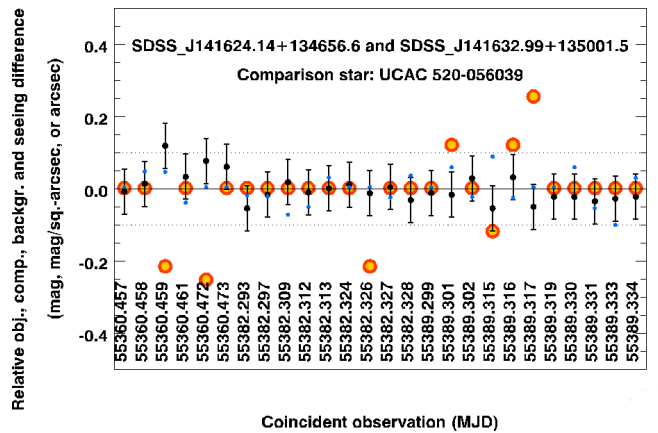}
\caption{Differences in object, seeing, sky background and extinction for two particular quasars falling inside the same frame, spanning one month. These coincident observations demonstrate the good data quality and photometric accuracy obtained for the sample.}\label{figure_differences_special}
\end{figure}

Some fully coincident observations did occur, that is, object pairs with fractional overlap of unity (310 times). One such case with multiple repeated observations is shown in Figure~\ref{figure_differences_special}, for quasars SDSS J141624.14+134656.6 and SDSS J141632.99+135001.5, observed with GMOS-North 26 times together over the course of a month (the MJD observation times are indicated), when those two happened to fall inside the same frame. In most cases, the comparison star was the same (UCAC4 520-056039) and so there is no difference in measured sky background (red circle), extinction (yellow) or seeing (blue dots). Those differ, however, in the few cases where a different calibration star was found (automatically) for the other. Spanning these observations, the uniformity of the photometry is remarkably stable. It is evident that photometry of quasars near the mean sample brightness (catalog magnitudes of 19.7 mag and 19.8 mag) was carried out over multiple observations over month-long timescales accurate to 0.10 mag, which is considered the limit of significant measurable differences for the rest of the sample.

\begin{figure}
\plotone{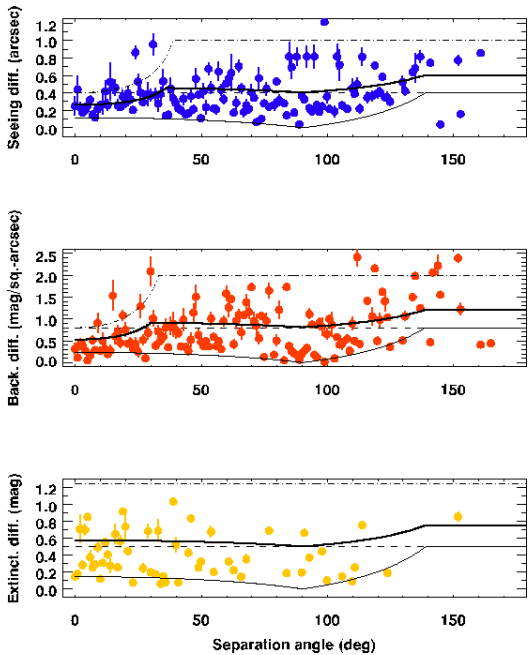}
\caption{Differences in seeing, sky background and extinction occurring over a single day; a nightly window starting at twilight at Cerro Pachon and lasting approximately 17 hours. Data are averaged in 1-degree-wide bins, with $1\sigma$ error bars.}\label{figure_differences_one_day}
\end{figure}

\begin{table*}[t]
\centering
\begin{tabular}{lrrlccrccr}
\hline
\hline
                             &R.A.            &Dec.            &Redshift    &Cat.       &Obj.               &Diff.      &Samp.    &Start           &Span\\
Name                         &(deg)           &(deg)           &$z$         &(mag)      &(mag)              &(mag)      &$n$      &(MJD)           &(days)\\
\hline
SDSS J090752.02+135829.2     &136.96675       & 13.97480       &1.7         &20.2       &$19.19\pm0.41$     &-1.01      &12       &56040.040       &   0.0239\\
SDSS J081839.27+574750.6     &124.66362       & 57.79739       &1.632       &20.6       &$19.85\pm0.12$     &-0.75      &12       &55153.474       &   0.0241\\
SDSS J081838.61+580235.7     &124.66088       & 58.04328       &1.2         &21.0       &$20.64\pm0.07$     &-0.36      &12       &55158.523       &   0.0248\\
SDSS J120835.93+020559.4     &182.14975       &  2.09986       &2.302       &20.2       &$20.35\pm0.06$     & 0.15      &10       &55270.223       &   0.0267\\
SDSS J120843.80+020840.8     & 82.18250       &  2.14467       &1.0         &19.6       &$20.46\pm0.06$     & 0.86      &10       &55270.223       &   0.0267\\
SDSS J111010.37+011423.8     &167.54321       &  1.23994       &0.4         &22.0       &$20.85\pm0.72$     &-1.15      &14       &55587.294       &   0.0460\\
SDSS J002251.40+155652.3     &  5.71417       & 15.94789       &2.235       &20.4       &$19.85\pm0.17$     &-0.55      &17       &55774.540       &   0.0702\\
3C 186.0                     &116.07279       & 37.88811       &1.068634    &17.5       &$17.10\pm0.11$     &-0.40      &11       &54151.217       &   0.1140\\
SDSS J042619.41+165726.8     & 66.58092       & 16.95747       &0.5         &20.4       &$18.65\pm0.13$     &-1.75      &30       &52591.485       &   1.0454\\
2QZ J112636.9+003454         &171.65408       &  0.58208       &0.550493    &17.9       &$18.33\pm0.30$     & 0.43      &10       &54918.397       &   1.0872\\
MC 1043-291                  &161.41917       &-29.45722       &2.128       &18.9       &$18.45\pm0.21$     &-0.45      &18       &53795.325       &   1.8113\\
NBCK J140849.77-010850.5     &212.20737       & -1.14739       &2.0         &21.4       &$20.80\pm0.40$     &-0.59      &29       &53848.160       &   2.0415\\
SDSS J135335.92+401723.1     &208.39971       & 40.28975       &1.9         &20.9       &$20.05\pm0.16$     &-0.85      &23       &52701.576       &   2.0511\\
SDSS J003027.98+261804.2     &  7.61658       & 26.30119       &1.534       &20.1       &$19.88\pm0.05$     &-0.22      &10       &55471.265       &   4.1237\\
VA-562                       &  7.60862       & 26.28028       &0.269       &18.2       &$18.92\pm0.05$     & 0.72      &22       &55471.263       &   4.1261\\
SDSS J095155.67+220947.5     &147.98200       & 22.16319       &0.634413    &17.9       &$17.67\pm0.12$     &-0.23      &10       &55296.298       &   9.9686\\
SDSS J095205.98+221018.8     &148.02492       & 22.17192       &2.627       &20.3       &$19.99\pm0.12$     &-0.31      &10       &55296.298       &   9.9686\\
SDSS J141624.14+134656.6     &214.10063       & 13.78242       &2.259       &19.8       &$20.88\pm0.07$     & 1.08      &27       &55360.457       &  28.8773\\
SDSS J141632.99+135001.5     &214.13750       & 13.83375       &1.0         &19.7       &$19.93\pm0.09$     & 0.22      &29       &55360.457       &  28.8773\\
SDSS J023639.93+282308.2     & 39.16642       & 28.38561       &1.9         &21.1       &$20.12\pm0.07$     &-0.98      &12       &54707.543       &  35.9549\\
SDSS J023653.25+282142.3     & 39.22188       & 28.36178       &1.0         &19.9       &$19.24\pm0.06$     &-0.66      &12       &54707.543       &  35.9549\\
IXO 10                       & 50.66833       &-37.27778       &0.515       &19.5       &$20.04\pm0.36$     & 0.54      &15       &54760.314       & 308.0033\\
IXO 69                       &190.90213       & 11.50256       &1.195       &18.9       &$18.18\pm0.19$     &-0.72      &15       &54180.450       & 682.9031\\
LBQS 1308-0104               &197.83021       & -1.34192       &2.620       &17.5       &$18.18\pm0.20$     & 0.68      &11       &52267.626       & 730.0214\\
SDSS J002235.96+001850.0     &  5.64983       &  0.31390       &1.6         &20.1       &$20.62\pm0.27$     & 0.52      &23       &52141.458       &1383.9300\\
CXOMP J01527-1359            & 28.18250       &-13.98361       &0.821       &20.9       &$20.46\pm0.35$     &-0.44      &10       &52474.553       &2244.6849\\
\hline
\hline
\end{tabular}
\caption{Multiple Significant Object Differences Ordered by Span of Samples\label{table_special}}
\end{table*}

Several repeat, correlated observations having brightness differences larger than the significance limit are provided in Table 1, ordered by the span of observations. Only those with at least 10 occurrences are retained; uncertainties reported are standard deviations. These are especially correlated; the most in the study. They were observed from both North and South, and through the full range of catalog brightness. There are notably cases of bright (up to 17.5 mag) quasars, although none is reported in the literature as a known variable. For example, 3C 186.0 ($z=1.07$) is a well-studied radio-loud source with a prominent jet \cite{Chiaberge2017}: observed 11 times beyond the significance limit, on average $0.40$ mag less than its catalog Red magnitude (and even brighter than its SDSS r value of 17.88). The largest discrepancies from the catalog are over a magnitude, occurring in fainter objects, as expected.  Monitoring of targets might reveal intrinsic, intra-day brightness fluctuations. There was, however, no case of significant object difference between the two telescopes obtained during a single-day window. Even so, those observations with one object in the FoV (either North or South; not both), are shown in Figure~\ref{figure_differences_one_day}.  There are some outliers, for instance, occasions of particularly good seeing at low elevation. But it can be seen how expectations of intra-day variation in seeing, sky background, and extinction values - for those cases that had a comparison star suffering less than 2 mag of extinction - are consistent with observations.

\begin{figure*}
\plotonewide{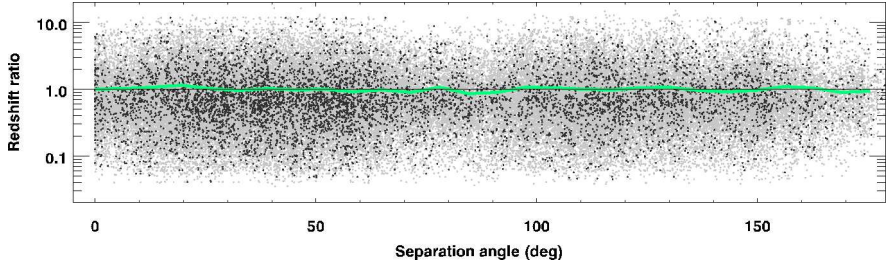}
\caption{Ratios of redshifts between objects ordered by separation angle, for all cross-correlated observations (grey points) and those above significance (black); averages in six-degree-wide bins (thick, green), close to an expected perfect median of unity.}\label{figure_fraction_with_angle}
\end{figure*}

Full cross correlation of the entire catalog indicates no evident bias in sample selection. Figure~\ref{figure_fraction_with_angle} shows the ratio of redshifts of any two objects as a function of angular separation. Although the visibility of objects from the two sites necessarily results in density variation of this distribution, it is fair, with no average redshift difference across the sample (thick, green). Ideally, the test sample would include only $z \geq 3.65$ objects $180^{\circ}$ apart, but with just 274 such observations taken instead across the full sky, there was not a sufficient sample to provide a meaningful comparison of just those. No observation of truly acausal pairs was obtained.

\begin{figure}
\plotonefull{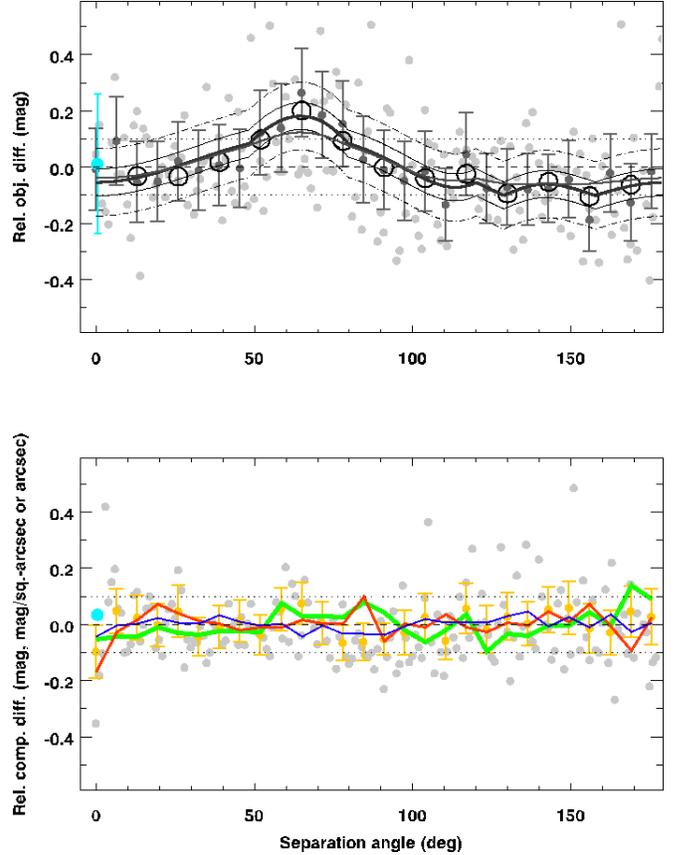}
\caption{Object (top) and comparison-star differences (bottom) per angular degree (light grey) and, for objects, averaged in 6-degree-wide bins (dark grey) within 68\%-confidence error bars, relative to the overall mean (dashed). Light-blue filled circles indicate the coincident cases only. Bottom panel: binned results for comparison stars (yellow); the standard deviation of cross-correlated differences is 0.024 mag. Compared to the significance limit (dotted) the differences in catalog magnitudes are flatter (thick, green) as are sky brightness (red) and image quality (thin, blue).  Top panel: signal is consistent with correlation as predicted by equation 13 (thick black curve) and open black circles show averages over two optimal angular bins, falling within twice the zeropoint error of that curve (thin black curves), thus reaching the sensitivity limits of the photometry; for comparison, the average of $2\times 10^6$ random trials (thin grey curve) and $5\times$ the threshold-limits for residual correlation (dot-dashed) are also shown.}\label{figure_with_angle_final}
\end{figure}

Results are shown in the top panel of Figure~\ref{figure_with_angle_final}: the most interesting being that for the full cross-correlated sample there is significant object correlation, even if that does not directly impact a true loophole-free test. Only cases where objects were within 2 mag of each other and having a relative Blue-Red colour difference less than 0.20 mag were retained. These are displayed as the relative differences in the objects, averaged in 1-degree-wide bins (grey) and 6-degree-wide bins (black). The error bars are their $1\sigma$ standard deviations. The results of just the coincident pairs are shown in light blue, that is, those cases where two objects fell inside the FoV. Displayed in the bottom panel are the differences of comparison stars, for reference; seeing (thin, blue) sky brightness (red), and the catalog magnitudes (thick, green) are overplotted. These are all relatively flat (the standard deviation in comparison magnitude differences is 0.024 mag), which makes the comparison to equation 13 (top panel: thick black curve) remarkable, in agreement at all optimally-binned angles and strongly against flatness (dashed), for the deviation expected, at $S/N=3$. It should be emphasized that this is not a fit, and all parameters were specified in Section~\ref{introduction} by the sample conditions. But analysis of the cumulative distribution functions do confirm a good match: the Anderson-Darling (AD) test statistic is under 0.004, and it rejects the null-hypothesis of flatness at 97.7\% probability (${\rm AD}=0.886$, p-value of 0.023). Results are similar, with expectedly more scatter, if only ${\rm r}$-band frames are used.

\begin{figure}
\plotonetiny{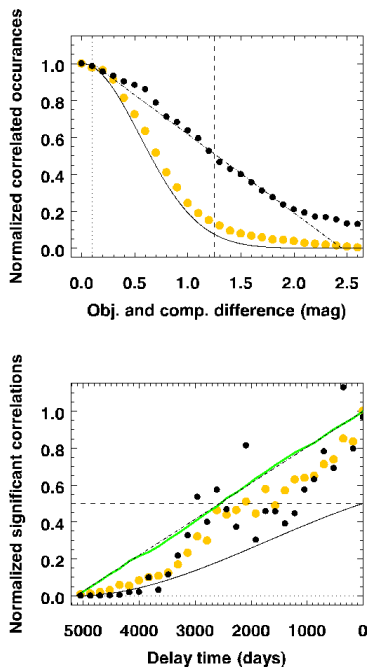}
\caption{Top panel: normalized occurrences of differences between objects (black) and comparisons (yellow).  Bottom panel: same for significant differences only, as a function of delay time between correlated occurrences in days; thick green curve is the result when the order of data is randomized.}\label{figure_significant}
\end{figure}

These deviations of differences in object brightnesses are significant relative to the measured errors; they cannot be accounted for by photometric uncertainty. A significance limit of 0.10 mag is indicated by a horizontal dotted line in Figure~\ref{figure_with_angle_final}. The enhancement as per equation 13 near $65^{\circ}$ is secure; structure beyond $120^{\circ}$ perhaps less so, but still consistent with the data. The distribution of these differences is displayed as well in Figure~\ref{figure_significant}. In the top panel a vertical dotted line shows the limit of 0.10 mag; a thin black curve is a Gaussian of width 0.25 mag, consistent with purely photometric error; a dashed line is the expected difference for a uniform sample of width 1.25 mag. This is intuitively the limit one would expect, if the sample was randomly drawn from the same distribution, 1.00 mag wide.  Note the four or five instances near differences of about 0.50 mag that are above this line, so occurring slightly more often than one would expect from a randomly drawn sample.

With reversed seasons between North and South hemispheres, one would expect some seasonal dependence on when objects were observed, and so although they may be selected on the chance they appear in either telescope FoV this selection will not be uniform over the year. The bottom panel averages in 6-month-long bins those cases which were beyond the significance limit from the sample (black). The thin black curve shows the result if half the sample were selected, decaying geometrically back towards the beginning of the window. The yellow circles are the comparisons for the full sample, the thick green line is the same except sample times have been randomized across the entire 14.5 years. Note that this resampling has no effect on the results displayed in Figure~\ref{figure_with_angle_final}; those are absolute values, so the order in which the differences were taken is not relevant, and those have already been cross correlated for the full sample. It is interesting that there appear to be some periods when it was more probable than random that there would be significant differences between two objects. The long timescales of those suggests that conducting an experiment avoiding such a bias may require repeat samples spanning years.

\section{Summary and Future Directions}\label{summary}

A clean QM experiment employing antipodal quasars requires that the measured fluxes of objects at both far-separated telescopes are above the local noise, and found uncorrelated. If not, this could hide synchronization; a potential case has been described, generated by averaging over all experimental geometries allowed by an Earth-wide test. An implication is that to be successful at triggering-source antipodes, some residual correlation should be found with lesser angles. Fair, random sampling across the sky might sense this signal, if not photon-by-photon, at least on the timescale of minutes. By mining the full GMOS-North and GMOS-South archive, approximately 30,000 broadband ${\rm g}$, ${\rm r}$, ${\rm i}$ and ${\rm z}$ quasar images were found, with many repeated; the vast majority of which are merely serendipitous: targeting an unrelated object in the field.  Photometry of each frame, which includes a stellar calibration for uniform data-quality selection, allows a careful analysis of the impact of source colour, variable sky conditions and airmass (sky background, extinction, and image quality) without explicit target or observer bias.  Almost 10000 sufficiently deep observations for Gemini GMOS North and South, complete with a nearby unsaturated UCAC4 star, allowed photometry with a global zeropoint uncertainty of about 2\% over a sample spanning 14.5 years. A ``virtual test" was performed on those data, comprising roughly 2 million observational pairs, which in their aggregate have 0.25-mag $1\sigma$ uncertainty within 6-degree-wide bins. This is sufficient to show an expected lack of flatness in relative object flux-differences with angular separation. A residual connection between distant sources is admittedly a somewhat surprising result, but not ruled out: the mean separation and redshift of those are less than could cause a conflict with causality, nor does it obviate previous QM tests. Others have recently considered the possibility of finding primordial correlations at even far-earlier epochs, for example, in the Cosmic Background Radiation \citep{Chen2019}.

Although the Gemini sample covers the full sky and range of possible angular separations between objects, there were not sufficient $z\geq3.65$ sources to provide a meaningful comparison restricted to those. A potentially confusing factor may be shifting bandpasses with redshift, and correcting to a common colour; a better technique may be spectroscopic, focused on bright emission lines. There was also limited information on how those individual sources (or the calibration stars) may have varied during this time. A subset of the data with significant differences is one output of this work, and provides a baseline from which to compare. Repeat observations of these at higher photometric precision would seem to provide a check on either real, intrinsic correlation between those source fluxes or false, spurious correlation due to unidentified source-selection or instrumental effects.  Those were controlled here by the telescopes and instruments being essentially identical, and blind selection from a prior independent catalog, but a wealth of archival sources of photometry from other telescopes could be added together to improve on this result too; multiple cross-calibration may actually serve to reduce zeropoint errors. Future facilities allowing precision long-term monitoring, such as with the Vera C. Rubin Observatory (previously referred to as the Large Synoptic Survey Telescope) will make false correlation via these potential error modes much easier to rule out. 

The current dataset provides only a small number of truly coincident observations (yielding just those cases where two quasars were in the same GMOS FoV) yet that is the goal of this endeavour. Note that fair, unbiased switching in an Earth-wide QM experiment would depend on simultaneous control of local noise sources, however the discriminators are set.  No such apparatus was implemented here, nor is a new experimental protocol suggested apart from this key improvement: employing widely-separated optical observatories allows sensing the apparent bias seen in the data, which is impossible for the constricted angular separations available to any single site. An attractive aspect of Gemini is that these are at two premier sites over 10600 km apart, with a combined view stretching a full $180^{\circ}$ across the sky. Ultimately, the coming era of 30-metre telescopes in both hemispheres is anticipated, with a $d^2$ aperture-diameter advantage of about $(30/8)^2=14$, plus a smaller PSF (and sky-background error) with adaptive optics increasing that to $d^4\sim 200$, bringing exposure times for $z \approx 4$ quasars down to a second, not minutes. Thus, low-noise, truly-synchronous photometry could sample timescales (in the quasar restframe) shorter than the round-trip light travel time between telescopes, and so unambiguously exclude any collusion between measurements due to local noise. As neither the emission processes at either external source nor switching-decision at either telescope could have influenced the other, the results would have to be pre-determined before the photons left the quasars. In short, the experimental outcome would imply a ``cosmic conspiracy" dating back nearly $90\%$ of the look-back time for the visible Universe.

\acknowledgments{I gratefully acknowledge many thoughtful comments on QM experiments by anonymous reviewers of earlier manuscripts, which helped improve on the original text. David Kaiser helpfully directed me to recent quasar tests of Bell's theorem and associated measurement criteria. This research used the facilities of the Canadian Astronomy Data Centre operated by the National Research Council of Canada with the support of the Canadian Space Agency, and I am especially grateful to David Bohlender for kind assistance with TAP scripts.}


\begin{thebibliography}{17}

\bibitem[Einstein, Podolsky and Rosen(1935)]{Einstein1935} A. Einstein, B. Podolsky \& N. Rosen, ``Can Quantum-Mechanical Description of Physical Reality Be Considered Complete?" Phys. Rev. {\bf 47}, 777 (1935)

\bibitem[Bell(1964)]{Bell1964} J.S. Bell, ``On the Einstein Podolsky Rosen Paradox," Physics Vol. 1, No. 3, 195, (1964)

\bibitem[Rosenfeld et al.(2017)]{Rosenfeld2017} W. Rosenfeld, D. Burchardt, R. Garthoff, K. Redeker, N. Ortegel, M. Rau \& H. Weinfurter, \prl {\bf 119}, 010402 (2017)

\bibitem[Friedman, Kaiser and Gallicchio(2013)]{Friedman2013} A.S. Friedman, D.I. Kaiser \& J. Gallicchio, \prd {\bf 88}, 044038 (2013)

\bibitem[Gallicchio, Friedman and Kaiser(2014)]{Gallicchio2014} J. Gallicchio, A.S. Friedman \& D.I. Kaiser, \prl {\bf 112}, 110405 (2014)

\bibitem[Handsteiner et al.(2017)]{Handsteiner2017} J. Handsteiner, A.S. Friedman, D. Rauch, J. Gallicchio, B. Liu, H. Hosp, J. Kofler, D. Bricher, M. Fink, C. Leung, A. Mark, T. Hien, H.T. Nguyen, I. Sanders, F. Steinlechner, R. Ursin, S. Wengerowsky, A.H. Guth, D.I. Kaiser, T. Scheidl \& A. Zeilinger, \prl {\bf 118}, 060401 (2017)

\bibitem[Li et al.(2018)]{Li2018} M.-H. Li, C. Wu, Y. Zhang {\it et al.}, \prl {\bf 121}, 080404 (2018)

\bibitem[Wu et al. (2017)]{Wu2017} C. Wu {\it et al.}, \prl 118, 140402 (2017)

\bibitem[Leung(2018)]{Leung2018} C. Leung, A. Brown, H. Nguyen, A.S. Friedman, D.I. Kaiser \& J. Gallicchio, \pra {\bf 97}, 042120 (2018)

\bibitem[Rauch et al.(2018)]{Rauch2018} D. Rauch, J. Handsteiner, A. Hochrainer {\it et al.}, \prl {\bf 121}, 080403 (2018)

\bibitem[Clauser et al.(1969)]{Clauser1969} J.F. Clauser, M.A. Horne, A. Shimony \& R.A. Holt, \prl {\bf 23}, 880 (1969)

\bibitem[MacLeod et al.(2010)]{MacLeod2010} C.L. MacLeod, Ž. Ivezić, C.S. Kochanek, S. Kozłlowski, B. Kelly, E. Bullock, A. Kimball, B. Sesar, D. Westman, K. Brooks, R. Gibson, A.C. Becker \& W.H. de Vries, \apj {\bf 721}, 1014 (2010) 

\bibitem[Mudd et al.(2018)]{Mudd2018} D. Mudd {\it et al.}, \apj {\bf 862}, 123 (2018)

\bibitem[Friedman et al.(2019)]{Friedman2019} A.S. Friedman, A.H. Guth, M.J.W. Hall, D.I. Kaiser \& J. Gallichio, \pra {\bf 99}, 012121 (2019)

\bibitem[Kasten(1965)]{Kasten1965} F. Kasten, ``A New Table Table and Approximation Formula for the Relative Optical Air Mass" (1965)

\bibitem[Chiaberge et al.(2017)]{Chiaberge2017} M. Chiaberge, J.C. Ely, E.T. Meyer, M. Georganopoulos, A. Marinucci, S. Bianchi, G.R. Tremblay, B. Hilbert, J.P. Kotyla, A. Capetti, S.A. Baum, F.D. Macchetto, G. Miley, C.P. O’Dea, E.S. Perlman, W.B. Sparks \& C. Norman, C., Astron. \& Astrophys. {\bf 600}, 57 (2017)

\bibitem[Chen et al.(2019)]{Chen2019} J.-W. Chen, S.-H. Dai, D. Maity, S. Sun \& Y.-L. Zhang, \prd {\bf 99}, 023507 (2019)

\end{thebibliography}
\end{document}